\documentclass[aps,twocolumn,prb,amsmath,amssymb,showpacs,superscriptaddress,floatfix]{revtex4-1}
\usepackage{bm}
\usepackage{epsfig}
\usepackage[usenames]{color}
\usepackage{graphicx}
\usepackage[hypertex]{hyperref}
\usepackage{esint}

\newcommand{\beq}{\begin{equation}}
\newcommand{\eeq}{\end{equation}}
\newcommand{\beqa}{\begin{eqnarray}}
\newcommand{\eeqa}{\end{eqnarray}}

\newcommand{\Em}{\mathcal E}

\DeclareMathOperator{\Imm}{Im}

\DeclareMathOperator{\Rem}{Re}

\def\XXint#1#2#3{{\setbox0=\hbox{$#1{#2#3}{\int}$ }
\vcenter{\hbox{$#2#3$ }}\kern-.5\wd0}}

\newcommand*{\PVI}{\fint}

\begin{document}
\date{\today }

\title{Pseudogap and zero-bias anomaly due to 
fluctuation suppression of quasiparticle tunneling}

\author{A.~Glatz}
\affiliation{Materials Science Division, Argonne National Laboratory, 9700 S.Cass Avenue, Argonne, Illinois 60637, USA}
\affiliation{Department of Physics, Northern Illinois University, DeKalb, Illinois 60115, USA}

\author{A.~A.~Varlamov}
\affiliation{Materials Science Division, Argonne National Laboratory, 9700 S.Cass Avenue, Argonne, Illinois 60637, USA}
\affiliation{CNR-SPIN, Viale del Politecnico 1, I-00133 Rome, Italy}

\author{V.~M.~Vinokur}
\affiliation{Materials Science Division, Argonne National Laboratory, 9700 S.Cass Avenue, Argonne, Illinois 60637, USA}

\begin{abstract}
We study the effect of superconducting fluctuations on the tunnel current-voltage
characteristics of disordered superconducting films placed in a perpendicular
magnetic field, $H$, in the whole $H$-$T$ phase diagram outside the superconducting region.
This tunnel-current is experimentally accessible by STM measurements.
In the domain of temperatures $T\geq T_{c0}$ and relatively weak fields $
H\ll H_{c2}(0)$ we reproduce existing results for the zero-voltage tunneling
conductance, but also discover an important nonlinear contribution, which
appears due to dynamic fluctuation modes and results in the
formation of a strong zero-bias anomaly (ZBA) on the scale $eV\sim k_{%
\mathrm{B}}(T-T_{c0})$.
At large voltages ($eV\sim k_{\mathrm{B}%
}T_{c0}$) these modes, together with the contribution from static fluctuations, form a \textit{pseudogap} maximum.
At low temperatures, with  magnetic field values near $H_{c2}(0)$, fluctuations acquire quantum
character and the general picture of the voltage dependent tunneling conductance
resembles that one close to $T_{c0}$, where the role of temperature and
magnetic field are exchanged.
In particular, a gap-like structure appears with maximum at $eV_{\max
}\sim \Delta _{\mathrm{BCS}}$ and a sharp ZBA on the scale $eV\sim \Delta
_{\mathrm{BCS}}(H/H_{c2}(0)-1)$.
The complete expression for the tunneling current at arbitrary fields and temperatures can be evaluated only numerically, which is presented in detail.
\end{abstract}


\maketitle

\section{Introduction}
According to the microscopic BCS theory~\cite{BCS}, the superconducting state is characterized by a gap in the normal excitation spectrum, centered at the Fermi level, $E_F$, which vanishes along the transition line $H_{c2}(T)$.
However, it was predicted, as early as in 1970~\cite{ARW70}, that  even above the superconducting critical temperature, $T_{c0}$ in zero magnetic field, in its normal phase, thermal fluctuations result in a noticeable suppression of the density of states, $\nu \left( E\right)$, of the superconductor in an energy range around the Fermi level.
More specifically, in the case of a disordered thin film~\cite{CCRV90} the characteristic energy range of this
suppression is\cite{energy} $E_{0}\sim T-T_{c0}$ and its zero energy depth is given by $\delta \nu _{\left( 2\right) }^{\left( \mathrm{fl}\right)
}\left( E=0,T\right) =-0.1\mathrm{Gi}_{\left( 2\right) }\nu _{n}\left[
T_{c0}/\left( T-T_{c0}\right) \right] ^{2}$, where $\mathrm{Gi}_{\left(
2\right) }=1.3/(p_{F}^{2}ld)$ is the Ginzburg-Levanyuk number
characterizing the strength of fluctuations in the film, and $\nu _{n}$ is the density of the states of a normal metal at the Fermi level.

With the discovery of high temperature superconductivity, the investigation of the suppression of the density of states
and the domain of parameters, where it is observed in cuprate superconductors developed into a major research area.
It became known as the {\it pseudogap} state and is believed to be a key mechanism for high temperature superconductivity itself\cite{micklitz+prb09}.
A recent experimental report of the pseudogap in a conventional 2D superconductor,
a disordered ultrathin titanium nitride (TiN) film~\cite{Benjamen10}, added another piece of the pseudogap puzzle.
Combined scanning tunneling microscope (STM) and transport measurements of the $I$-$V$
characteristics allowed to carefully compare these experimental findings with theoretical predictions
of the behavior of the contribution to transport conductivity due to superconducting fluctuations (SF)~[\onlinecite{VD83}].
As a result, the observed pseudogap, existing over a wide range of temperatures above $T_{c0}$, was attributed
to two-dimensional (2D) SF, which are favored by the proximity to the superconductor-insulator transition~\cite{Benjamen10}.
This recent finding paralleled an earlier observation of the pseudogap due to SF in experiments on Al-I-Al tunnel junctions~\cite{Kh84} that have been observed soon after its prediction and poses a call to re-visit the pseudogap state in high-$T_c$ cuprates experimentally and carefully inspect the role of superconducting fluctuations.

A major experimental tool for determining the density of states is by measurements of the differential tunnel conductivity
\begin{align}
\sigma _{\mathrm{tun}}\left( V\right) & =dI\left( V\right) /dV  \notag \\
& \sim \int_{-\infty }^{\infty }\left( -\frac{\partial n_{F}\left(
E,T\right) }{\partial E}\right) \nu \left( E+eV\right) dE\,.  \label{eq.sigmatun}
\end{align}%
Considering that the success in revealing the nature of the pseudogap in superconducting films~\cite{Benjamen10}
was based on the ability to identify the effect of SF on the electronic transport, one immediately recognizes the quest
for uncovering the role of fluctuations in tunneling properties.
The knowledge of the behavior of the fluctuation contribution to $\sigma _{\mathrm{tun}}$
would have offered an irreplaceable technique for identifying fluctuation effects directly from the experimental data.
Our work addresses this challenging task and develops a theory for the fluctuation contribution
to the tunneling conductance.

\section{Model}
We study the effect of SF on the tunneling current $I\left(V\right) $ between a normal metal electrode and a disordered two-dimensional superconducting film placed in perpendicular magnetic field throughout the whole phase diagram above the $H_{c2}(T)$ line.
Describing this system by means of a tunnel Hamiltonian, the tunnel-current can be expressed in terms of the correlator $K\left( \omega _{\nu }\right) $ of the electron Green's functions of the corresponding electrodes, which is analytically continued from Matsubara frequencies $\omega _{\nu }=2\pi T\nu ,$ $\nu =0,1,2,...$ to the upper half-plane of complex frequencies $\omega _{\nu}\rightarrow -i\omega =-ieV$,~\cite{VD83}:
\begin{equation}
\sigma\left( V\right) =-e\Imm K^{R}(eV).  \label{IKgen}
\end{equation}

\begin{figure}[tbh]
\begin{center}
\includegraphics[ width=0.7\columnwidth ]{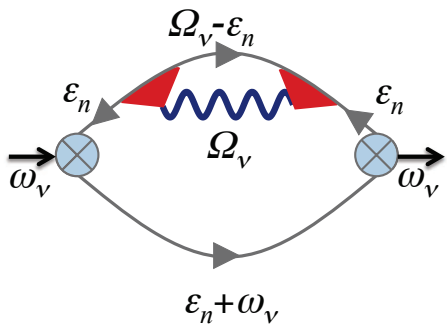}
\end{center}
\caption{(Color online) Diagram contributing to the fluctuation tunnel-current
in first orders of barrier transparency and strength of fluctuations $%
\mathrm{Gi}_{\left( 2\right) }$. Solid lines correspond to single-electron
Green's functions of the electrodes, the wavy line represents the fluctuation
propagator, crossed circles correspond to matrix elements of the tunneling
Hamiltonian, and the (red) solid triangles are Cooperon vertices accounting for impurity
averaging.}
\label{fig1}
\end{figure}

Being interested in low-transparency junctions and restricting our
consideration to the first order in $\mathrm{Gi}_{\left( 2\right) }$, one
can see that in the case the second electrode is not subject to superconducting fluctuations -- e.g., is a normal STM tip --
the only diagram which contributes to the tunnel-current is that presented in Fig.~\ref{fig1}.
This diagram describes the suppression of the tunnel-current due to the mechanism  of
fluctuation renormalization of the quasi-particle density of states, discussed above.

In the absence of magnetic fields, the correlation function, Eq.~(\ref{IKgen}), was already studied
in momentum representation~\cite{VD83}.
The generalization to the case of a perpendicular magnetic field can be made by going over from the
momentum to Landau representation with an appropriate quantization of the
Cooper pair motion (see, for example, Refs.~[\onlinecite{LV05, GVV11,glatz+epl11}]).
Formally, this corresponds to a replacement of the energy associated with the motion of the center of mass of a free Cooper pair
with momentum $\mathbf{q}$ by the eigen-energy of the Landau state of level $m$: $\mathcal{D}\mathbf{q}^{2}$ $\rightarrow \omega_c\left( m+1/2\right)$.
Here $\mathcal{D}$ is the electron diffusion coefficient and $\omega_c=4e\mathcal{D}H$ is the cyclotron frequency corresponding to the rotation of the center of  mass of a Cooper pair in a magnetic field $H$.
The integration over the two-dimensional momentum in correlator~(\ref{IKgen}) is replaced by a
summation over Landau levels according to the rule:
\begin{equation*}
\frac{\mathcal{D}}{8T}\int \frac{d^{2}q}{(2\pi )^{2}}f\left[ \mathcal{D}q^{2}%
\right] =\frac{h}{2\pi ^{2}t}\sum_{m=0}^{M}f\left[ \omega _{\mathrm{c}}(m+%
\frac{1}{2})\right] ,
\end{equation*}%
where $M=(T_{c0}\tau)^{-1}$ is a cut-off parameter related to the elastic electron scattering time $\tau$ (see Ref.~[\onlinecite{GVV11}] for details).
This transformation is applied to the general expression for the correlation
function $K\left( \omega _{\nu }\right)$, and one finds \cite{VD83}:
\begin{align}
& K\left( \omega _{\nu }\right) =K^{\left( \mathrm{reg}\right) }\left(
\omega _{\nu }\right) +K^{\left( \mathrm{an}\right) }\left( \omega _{\nu
}\right)  \label{K1+K2} \\
& =\frac{2T_{c0}Sh}{\pi ^{3}\sigma _{n}R_{N}}\sum_{m=0}^{M}\left[
\sum_{k=0}^{\infty }+\sum_{k=-\nu }^{-1}\right] \frac{\left[ \Em%
_{m}^{\prime }\left( k+2\nu \right) -\Em_{m}^{\prime }\left(
k\right) \right] }{\Em_{m}\left( |k|\right) }  \notag
\end{align}%
with $\sigma _{n}=e^{2}\nu _{n}\mathcal{D},$ $R_{N}$ being the tunneling resistance
of the junction and $S$ is its surface area.
The function
\begin{equation}
\Em_{m}\left( x\right) =\ln t+\psi \left[ \frac{1+x}{2}+\frac{4h}{%
\pi ^{2}t}\left( m+\frac{1}{2}\right) \right] -\psi \left( \frac{1}{2}\right)
\label{em}
\end{equation}%
represents the denominator of the fluctuation propagator (wavy line
in Fig.~\ref{fig1}):
\begin{equation}
\mathcal{L}_{m}\left( x\right) =-\nu _{n}\Em_{m}^{-1}\left( x\right)
,  \label{LE}
\end{equation}%
written in Landau representation and describing the fluctuation pairing
of electrons in the normal phase of a superconductor over a wide range of
temperatures and fields~\cite{LV05}. Here $t=T/T_{c0}$ and $h=\pi ^{2}/(8\gamma _{E})
H/H_{c2}(0)$ are dimensionless temperature and magnetic field normalized by
critical temperature and the value of second critical field respectively, $\gamma _{E}=1.78$ is the exponential Euler constant.
The cyclotron frequency of a Cooper pair rotation in this parametrization is $ \omega _{\mathrm{c}}=\left(16hT_{c0}/\pi\right)$.
We clarify that $\Em_{m}^{\prime }\left( x\right)$ denotes derivative of the function
$\Em_{m}\left( x\right) $ with respect to its argument $x$, explicitly given by
\begin{equation}
\Em_{m}^{\prime }\left( x\right) =\frac{1}{2}\psi^{\prime } \left[
\frac{1+x}{2}+\frac{4h}{\pi ^{2}t}\left( m+\frac{1}{2}\right) \right] .
\label{deri}
\end{equation}

The two terms in Eq.~(\ref{K1+K2}) correspond to two fluctuation contributions to the tunnel-current
with different analytical properties.
Below we demonstrate how these contributions give rise to the pseudogap maxima and the ZBA singularity in the
tunneling conductivity in two-dimensional disordered superconductors.

\section{Complete expression for the fluctuation tunnel-current}

We start our analysis with the first term of Eq.~(\ref{K1+K2}).
Since the external frequency $\omega _{\nu }$ enters the expression for $K^{(%
\mathrm{reg})}\left( \omega _{\nu }\right) $ only via the argument of the
analytical function $\Em_{m}^{\prime }\left( k+2\nu \right) $ [see
Eq.~(\ref{K1+K2})], one can easily perform its analytical continuation just
by substitution $\omega _{\nu }\rightarrow -ieV$.
Using Eq.~(\ref{IKgen}), one finds for the general expression of the corresponding current $I^{(\mathrm{reg%
})}\left( V\right) $:
\begin{equation}
I^{(\mathrm{reg})}\left( V\right)\! =\!-\frac{2h}{\pi^{3}}\!\left( \frac {eT_{c0}S%
}{\sigma_{n}R_{N}}\right)\! \sum_{m=0}^{M}\sum_{k=0}^{\infty}\frac{\Imm%
\Em_{m}^{^{\prime}}\left( k\!-\!\frac{ieV}{\pi T}\right) }{\Em%
_{m}\left( k\right) }.  \label{I1}
\end{equation}

The second contribution to the tunneling current is determined by
\begin{equation}
K^{\left( \mathrm{an}\right) }\left( \omega _{\nu }\right) =\frac{2T_{c0}Sh}{%
\pi ^{3}\sigma _{n}R_{N}}\sum_{m=0}^{M}\sum_{k=1}^{\nu }f_{m}(k,\omega _{\nu
}),  \label{I2}
\end{equation}%
with%
\begin{equation}
f_{m}(k,\omega _{\nu })=\frac{\left[ \Em_{m}^{\prime }\left( 2\nu
-k\right) -\Em_{m}^{\prime }\left( k\right) \right] }{\Em%
_{m}\left( k\right) }.  \label{fm}
\end{equation}%
Here the analytical contribution is more complex than in the case of $K^{(%
\mathrm{reg})}\left( V\right) ,$ since the frequency\ $\omega _{\nu }$\ is
 not only present in the argument of function~(\ref{fm}) but also in the
upper limit of the sum over $k$ in Eq.~(\ref{I2}).
Note, that this summation limit can be reduced from $\nu $ to $\nu -1$ since $f_{m}(k=\nu ,\omega _{\nu })=0$.
The analytical continuation of a function of the form
\begin{equation*}
\theta _{m}\left( \omega _{\nu }\right) =\sum_{k=1}^{\nu -1}f_{m}(k,\omega
_{\nu })
\end{equation*}%
onto the upper half-plane of complex frequencies was performed in Ref.~[\onlinecite{AV80}] [see also Ref.~[\onlinecite{LV05}], equation (7.90)].
\begin{figure}[ptb]
\begin{center}
\includegraphics[ width=0.9\columnwidth ]{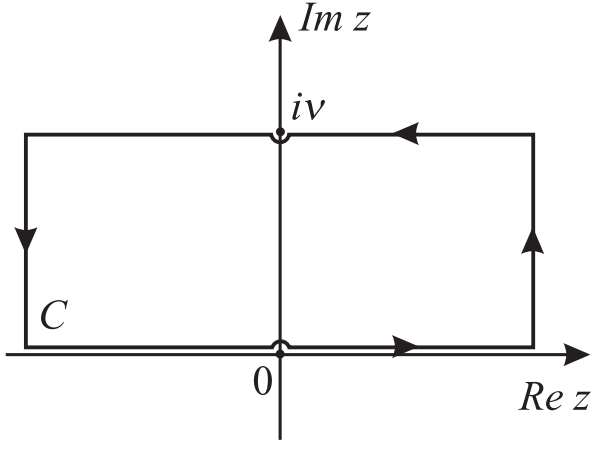}
\end{center}
\caption{Closed integration contour $\mathcal{C}$ in the plane of complex
frequencies.}
\label{mtcontour}
\end{figure}
By means of the Eliashberg transformation~\cite{E61} the corresponding sum can be presented as a counterclockwise integral over a closed contour $\mathcal{C}$ consisting of two horizontal lines, two vertical lines, and two semicircles in the upper complex plane, where the latter exclude the points $0$ and $i\nu$ (see Fig.~\ref{mtcontour}):
\begin{equation*}
\theta _{m}\left( \omega _{\nu }\right) \!=\!\frac{1}{2i}\ointctrclockwise\limits_{\mathcal{C}}%
\coth \left( \pi z\right) f_{m}(-iz,\omega _{\nu })dz\,.
\end{equation*}%
The integrals over the vertical line segments become zero, the integral over the semi-circle at $z=i\nu $ is zero since $f_{m}(k=\nu ,\omega _{\nu })=0$, the integral
over the semi-circle at $z=0$ reduces to the residual of $\coth \left( \pi z\right) $.
Inverting the direction of integration over the line segment with $\Imm z=\nu $ and then shifting the integration variable as $z+i\omega _{\nu }/2\pi T\rightarrow z_{1}$ in the corresponding integral, one finds:
\begin{eqnarray}
& \theta_{m}\left( \omega_{\nu}\right) \!=-\!\frac{f_{m}(0,\omega_{\nu})}{2}+%
\frac{1}{2i}\PVI_{-\infty}^{\infty}\coth\left( \pi z\right) \nonumber \\
& \times\left[ f_{m}(\!-iz,\omega_{\nu})\!-\!f_{m}(\!-iz\!-\!\omega_{\nu}/2%
\pi T,\omega_{\nu})\right] dz,  \label{resi}
\end{eqnarray}
where the ``dashed'' integral symbol means that the integral is performed in the sense of a Cauchy principal value.
Eq.~(\ref{resi}) is already an analytical function of $\omega _{\nu }$ and one can perform its continuation just by the standard substitution $\omega _{\nu }\rightarrow -i\omega$.
Shifting the variable in the second integral again as $z-\omega /2\pi T\rightarrow z_{2}$ and using the identity
\begin{equation*}
\coth a-\coth b=-\frac{\sinh \left( a-b\right) }{\sinh a\sinh b}
\end{equation*}%
one finally finds%
\begin{align}
& \theta _{m}^{R}\left( -i\omega \right) \!=-\!\frac{f_{m}(0,-i\omega )}{2}
\label{thetar} \\
& -i\frac{\sinh \left( \omega /2T\right) }{2}\PVI\limits_{-\infty
}^{\infty }\frac{f_{m}(-iz,-i\omega )dz\!}{\sinh \left( \pi z\right) \sinh
\pi \left( z+\omega /2\pi T\right) }.  \notag
\end{align}%
Substituting the explicit expression for function $f_{m}(-iz,-i\omega
)$ from Eq.~(\ref{fm}) into Eq.~(\ref{thetar}) results in
\begin{align}
& K^{\left( \mathrm{an}\right) R}\left( -i\omega \right) =-\frac{T_{c0}Sh}{%
\pi ^{3}\sigma _{n}R_{N}}\sum_{m=0}^{M}\left\{ \frac{\left[ \Em%
_{m}^{\prime }\left( -\frac{i\omega }{\pi T} \right) -\Em_{m}^{\prime }\left(
0\right) \right] }{\Em_{m}\left( 0\right) }\right.  \label{KR2} \\
& \left. +i\sinh \left( \frac{\omega }{2T}\right) \PVI\limits_{-\infty }^{\infty }%
\frac{\left[ \Em_{m}^{\prime }\left( iz\!-\!\frac{i\omega }{\pi T}\right)\!-\!
\Em_{m}^{\prime }\left(\!-iz\right) \right] dz\!}{\Em%
_{m}\left( -iz\right) \sinh \left( \pi z\right) \sinh \pi \left( z+\frac{\omega}
{2\pi T}\right) }\right\} .  \notag
\end{align}%

Eqs. (\ref{IKgen}) and (\ref{KR2}) determine the second fluctuation
contribution to the tunneling current $I^{\left( \mathrm{an}\right) }\left(
V\right)$.

Let us note that the first term of $K^{\left( \mathrm{an}\right) R}$ is
nothing but half of the first summand (with $k=0$) of the sum in Eq.~(\ref{I1}) with opposite sign.
Technically it would be easy to incorporate the latter into $K^{\left( \mathrm{reg}\right) R}$.  However, such a procedure would be physically misleading: we will see
below that this $k$ and $z$ independent term in Eq.~(\ref{KR2}) cancels the corresponding linear contribution stemming from the integral
term at small voltages.
As a result, the current $I^{\left( \mathrm{an}\right) }\left( V\right) $, determined by the imaginary part of Eqs.~(\ref{KR2}), does not contain a
linear contribution if expanded in powers of voltage. This means that it does not contribute to the magnitude of the
differential tunnel conductivity at zero voltage $\sigma _{\mathrm{tun}%
}^{\left( \mathrm{fl}\right) }\left( T,H,V=0\right) =dI^{\left( {\mathrm{fl}}%
\right) }/dV|_{V=0}$, which is the easiest quantity to measure in experiments.
Nevertheless, it contributes to the current-voltage characteristics at finite voltages and, as we will see below, can noticeably
manifest itself even at very low voltages $eV\sim T-T_{c0}$ as a ZBA.

Adding $I^{\left( \mathrm{reg}\right) }\left( V\right) $ and $I^{\left(
\mathrm{an}\right) }\left( V\right) $ one finds the general expression for the
fluctuation contribution to the tunnel-current, which is valid in the complete phase diagram beyond the $H_{c2}\left( T\right) $ line:
\begin{widetext}%
\begin{align}
&I^{\left(  \mathrm{fl}\right)  }\left(  t,h,V\right)  =I^{\left(
\mathrm{reg}\right)  }+I^{\left(  \mathrm{an}\right)  }=-\frac{2eT_{c0}Sh}%
{\pi^{3}\sigma_{n}R_{N}}\sum_{m=0}^{M}\sum_{k=0}^{\infty}\frac
{\Imm\Em_{m}^{\prime}\left(  k-ieV/\pi T\right)
}{\Em_{m}\left(  k\right)  }+\frac{eT_{c0}Sh}{\pi^{3}\sigma_{n}R_{N}%
}\sum_{m=0}^{M}\left\{  \frac{\Imm\Em_{m}^{\prime}\left(
-ieV/\pi T\right)  }{\Em_{m}\left(  0\right)  }\right.
\label{Itot}\\
&  \left.  +\sinh\left(  \frac{eV}{2T}\right)  \PVI\limits_{-\infty
}^{\infty}dz\frac{\operatorname{Re}\Em_{m}\left(  iz\right)  \left[
\operatorname{Re}\Em_{m}^{\prime}\left(  iz-ieV\right)
-\operatorname{Re}\Em_{m}^{\prime}\left(  iz\right)  \right]
+\Imm\Em_{m}\left(  iz\right)  \left[  \Imm%
\Em_{m}^{\prime}\left(  iz-ieV\right)  +\Imm%
\Em_{m}^{\prime}\left(  iz\right)  \right]  }{\sinh\left(  \pi
z\right)  \sinh\left[  \pi\left(  z-eV/2\pi T\right)  \right]  \left[
\Rem^{2}\Em_{m}\left(  iz\right)  +\Imm%
^{2}\Em_{m}\left(  iz\right)  \right]  }\right\}  .\nonumber
\end{align}
\end{widetext}

\begin{figure}[bth]
\begin{center}
\includegraphics[ width=0.99\columnwidth]{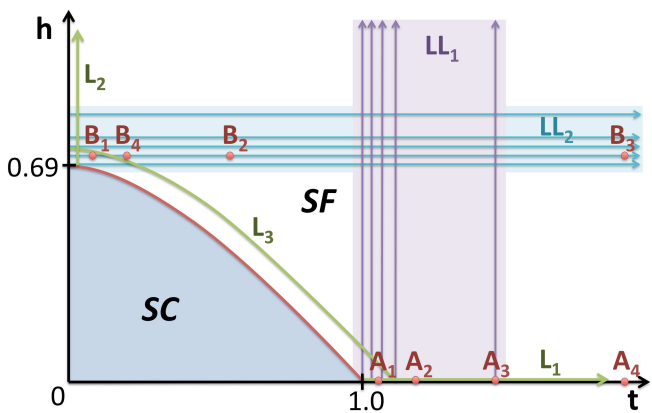}
\end{center}
\caption{(Color online)  Illustration of the $t$-$h$ phase diagram with the parameter space location of all plots in this paper. In each plot the parameter points ($A_i$ or $B_i$), parameter line cuts ($L_i$), or line groups ($\mathrm{LL}_i$) are given.}
\label{fig.regions}
\end{figure}

Eq. (\ref{Itot}) is the main result of this work.
The first term $I^{\left( \mathrm{reg}\right) }\left( V\right)$ has been studied in detail for different limiting cases using different approaches:
 close to $T_{c0}$,~\cite{VD83, CCRV90,V93,L10}: (i) in a wide temperature range in zero field~\cite{VD83}, or (ii) close to $T_{c0}$ in magnetic fields $H\ll
H_{c2}(0)$,~\cite{R93}.
The current contribution $I^{\left( \mathrm{an}\right) }\left(
V\right) $  has been omitted in all these works based on the ``standard''
argument that the zero frequency bosonic mode (which traverses through the propagator) is singular
in the vicinity of the transition.
However, it is known that this argument sometimes works (e.g., in the case of the Maki-Thompson contribution to conductivity~\cite{M68}),  but also sometimes fails (e.g., for the Aslamazov-Larkin contribution to conductivity~\cite{AL68}).
In our case this argument turns to out be correct only for very small voltages. The reason being that voltage itself, together with
temperature deviations from the transition point and finite magnetic fields, drives
the system away from the immediate vicinity of the transition, which invalidates the
argument regarding the dominance of the zero frequency bosonic mode.

\begin{figure*}[tbh]
\begin{center}
\includegraphics[ width=0.44\textwidth ]{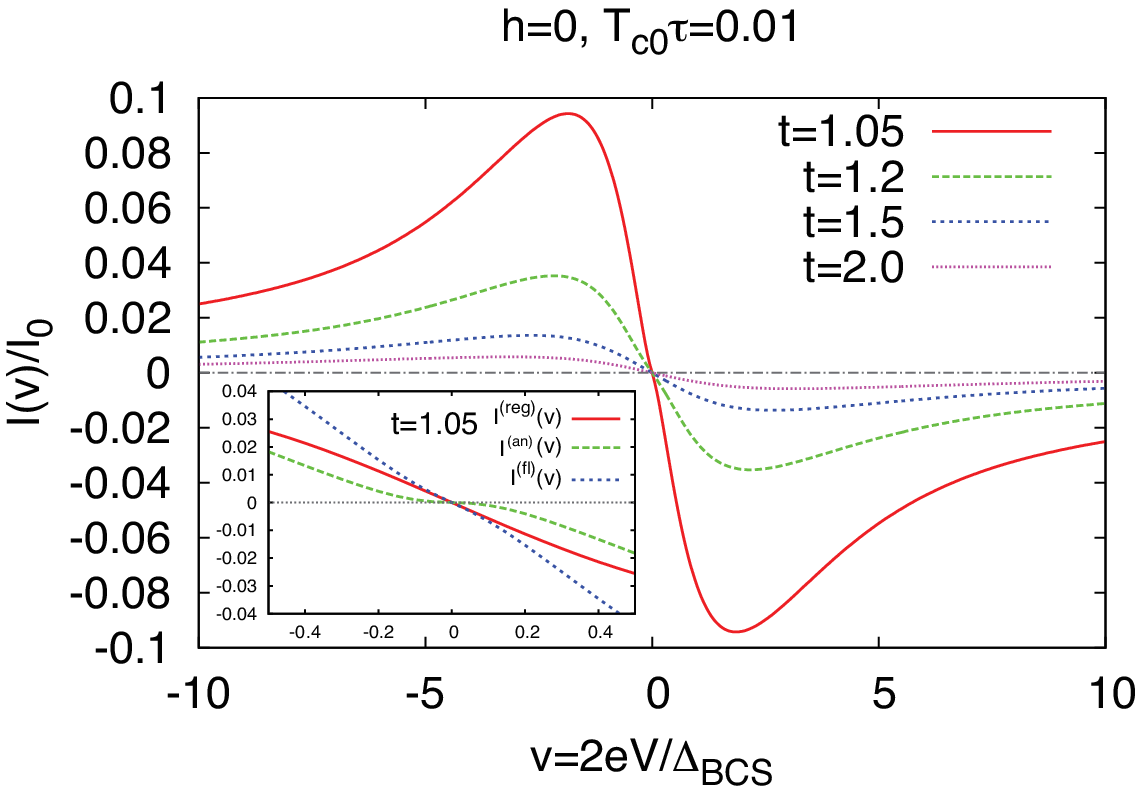}
\includegraphics[ width=0.44\textwidth ]{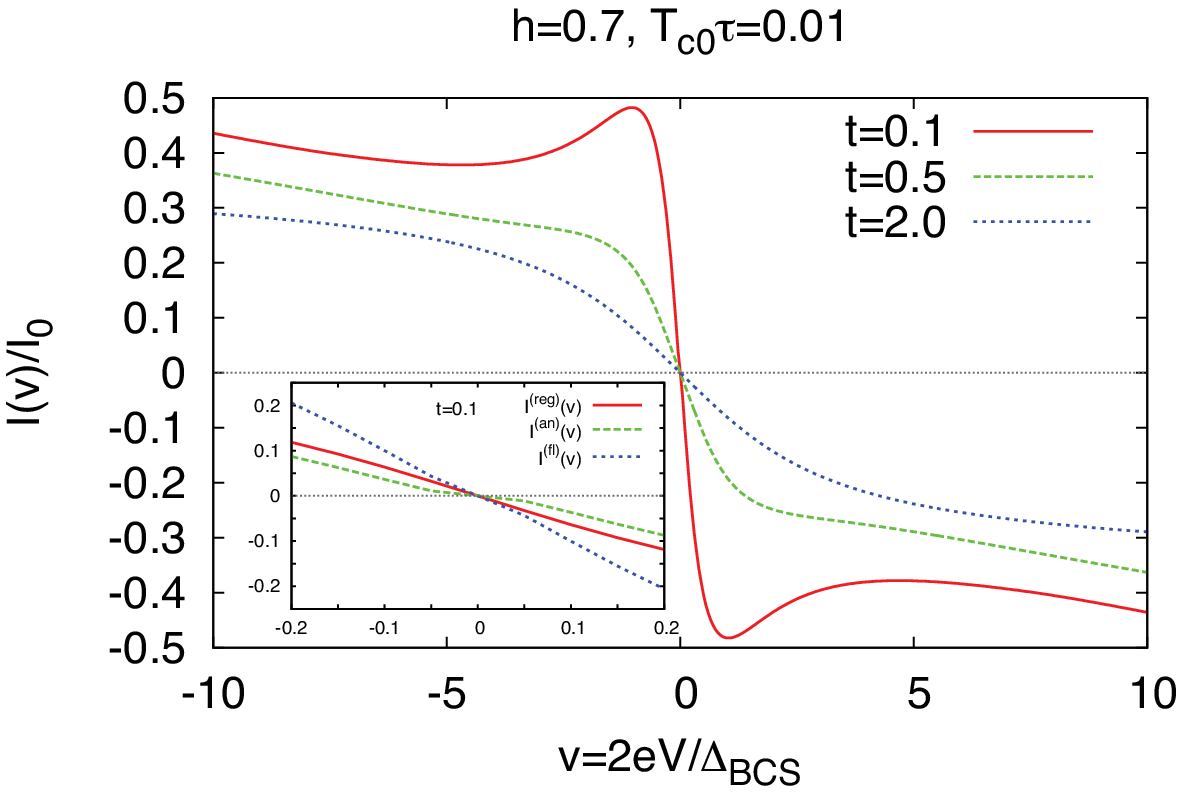}
\end{center}
\caption{(Color online)  Total tunneling current
close to $T_{c0}$ (left) and  near $h_{c2}(0)$ (right) at various temperatures depending on the dimensionless voltage $v=2eV/\Delta_{\mathrm{BCS}}$.
The insets show the regular and anomalous contributions at the respective lowest temperature separately. As one can see, the anomalous part has a nonlinear component near $v=0$. The current is normalized to $I_0=eT_{c0}S/(\sigma_n R_N)$. (left) parameter points in Fig.~\ref{fig.regions} are $A_1$-$A_4$, inset $A_1$, (right) parameter points  are $B_1$-$B_3$, inset $B_1$.}
\label{fig.tunnelIV}
\end{figure*}

We present several plots of the tunnel-current and the tunnel conductance. Since they depend on three parameters: $t$, $h$, and $v$, only lines or planes in the full parameter space are presented as line or surface plots.  We included Fig.~\ref{fig.regions}, showing all parameter points and lines in the $t$-$h$ phase diagram for all following figures.  The critical field line, $h_{c2}(t)$, separating the superconducting (SC) and the normal fluctuation region (SF) is defined by $\Em_0(0)=0$. Each figure caption refers to these parameter locations. 
Fig.~\ref{fig.tunnelIV} shows the behavior of $I^{\left(  \mathrm{fl}\right)  }\left(  t,h,V\right)$ near $T_{c0}$ and $H_{c2}(0)$.

In the following we will carefully analyze the effect of superconducting fluctuations in the
whole phase diagram. We start our discussion with the regular contribution $
I^{\left( \mathrm{reg}\right) }\left( V\right) $ and then elucidate the important role of the anomalous contribution $I^{\left( \mathrm{an}%
\right) }\left( V\right)$, which was neglected in literature so far.

\section{Analysis of  the asymptotic behavior of $I^{\left( \mathrm{reg}\right) }\left(
V\right) $}

\emph{Close to $T_{c0}$ and for sufficiently weak magnetic fields $H\ll
H_{c2}\left( 0\right) $}, the most singular term in Eq. (\ref{I1}) arises
from the zero frequency bosonic mode $k=0$, when the propagator has a pole at $\epsilon=0$
and
\begin{equation}
\Em_{m}\left( 0\right) =\epsilon +2h\left( m+\frac{1}{2}\right)
\label{em0}
\end{equation}%
with $\epsilon =\ln t\approx t-1\ll 1$ as reduced temperature. The summation
over Landau levels can be performed in terms of polygamma-functions, $\psi^{(n)}(x)$,  and one
finds an expression valid for any combination of $\epsilon$ and  $h\ll 1$:
\begin{align}
I^{(\mathrm{reg})}\left( V,h,\epsilon \right) =& -\frac{eTS}{2\pi ^{3}\sigma
_{n}R_{N}}\left[ \ln \frac{1}{2h}-\psi \left( \frac{1}{2}+\frac{\epsilon }{2h%
}\right) \right]   \notag \\
& \cdot \Imm\psi ^{\prime }\left( \frac{1}{2}-\frac{ieV}{2\pi T}\right).\label{eq.Ireg}
\end{align}%
Eq.~(\ref{eq.Ireg}) reproduces the results of Refs. ~[\onlinecite{VD83,R93}]. The corresponding contribution to the
tunneling conductance is
\begin{align}
\sigma ^{(\mathrm{reg})}\left( V\right) & =\frac{Se^{2}}{4\pi ^{4}\sigma
_{n}R_{N}}\left[ \ln \frac{1}{2h}-\psi \left( \frac{1}{2}+\frac{\epsilon}{%
2h}\right) \right]   \notag \\
& \cdot \Rem\psi ^{\prime \prime }\left( \frac{1}{2}-\frac{ieV}{2\pi T}%
\right) .
\end{align}

\emph{In the region of high temperatures $T\gg T_{c0}$ and zero magnetic
field} we restrict our analytical consideration to the
fluctuation contribution to the differential conductivity at zero voltage.
Performing an integration instead of a summation in Eq.~(\ref{I1}) one finds%
\begin{equation*}
\sigma ^{(\mathrm{reg})}(0,t\gg 1)=-\frac{Se^{2}}{4\pi ^{2}\sigma _{n}R_{N}}%
\left( \ln \frac{\ln \frac{1}{T_{c0}\tau }}{\ln t}\right) ,
\end{equation*}%
which is again in complete agreement with Ref.~[\onlinecite{VD83}].

\emph{Close to the line $H_{c2}\left( t\right) $ and for sufficiently low
temperatures $t\ll h_{c2}(t)$} the lowest Landau level approximation (LLL)
holds. The corresponding propagator (with quantum number $m=0)$ has
a pole structure and Eq.~(\ref{em}) acquires the form:
\begin{equation}
\Em_{0}\left( k\right) =\widetilde{h}+\frac{\pi ^{2}tk}{4h_{c2}}
\label{eline}
\end{equation}%
with $\widetilde{h}\left( t\right) =\left( H-H_{c2}\left( t\right) \right)
/H_{c2}\left( t\right)$. Keeping only the $m=0$ term in Eq. (\ref{I1}),
one can write
\begin{equation}
I^{(\mathrm{reg})}\left[V,t\!\ll\! h_{c2}(t)\right]\!=\!-\frac{2eT_{c0}Sh}{\pi
^{3}\sigma _{n}R_{N}}\sum_{k=0}^{\infty }\frac{\Imm\Em%
_{0}^{\prime }\left( k-\frac{ieV}{\pi T}\right) }{\widetilde{h}+\frac{\pi ^{2}tk}{4h_{c2}(t)}
}.  \label{sum4}
\end{equation}%
The imaginary part $\Imm\Em_{0}^{\prime }\left(
k-ieV/\pi T\right)$ can be explicitly written using Eq.~(\ref{deri})
in the limit $t\ll h_{c2}(t)$ and the asymptotic behavior of $%
\psi ^{\prime }\left( |x|\gg 1\right)\sim 1/x$:
\begin{equation}
\Imm\Em_{0}^{\prime }\left( k-\frac{ieV}{\pi T}\right) =\frac{eV%
}{2\pi T}\frac{1}{\left[ k+\frac{4h_{c2}(t)}{\pi ^{2}t}\right] ^{2}+\left( \frac{eV}{\pi T}\right) ^{2}}.
\end{equation}%
The summation in Eq.~(\ref{sum4}) can then be performed exactly in terms of
polygamma-functions, i.e., using
\begin{equation*}
\sum_{k=0}^{\infty }\frac{1}{k+\alpha }\frac{1}{\left( k+\beta \right)
^{2}+\gamma ^{2}}=\frac{1}{\gamma }\Imm\frac{\psi \left( \alpha \right)
-\psi \left( \beta +i\gamma \right) }{\beta +i\gamma -\alpha }\,,
\end{equation*}%
which gives an expression for the regular part of the fluctuation current valid for low enough temperatures along the line $h_{c2}(t)$:
\begin{widetext}
\begin{equation}
I^{\left( \mathrm{reg}\right) }\left[ v_{t},t\ll h_{c2}\left( t\right) %
\right] =-\frac{2eST_{c0}h}{\pi ^{3}\sigma _{n}R_{N}}\frac{v_{t}}{1+v_{t}^{2}%
}\left\{ \left[ \ln \left( \frac{4h_{c2}\left( t\right) }{\pi ^{2}t}\right)
\sqrt{1+v_{t}^{2}}\allowbreak \allowbreak -\psi \left( \frac{4h_{c2}}{\pi
^{2}t}\widetilde{h}\right) \right] -\frac{\arctan v_{t}}{v_{t}}\right\}.
\label{curr1a}
\end{equation}%
\end{widetext}

Here, we introduced the dimensionless voltage
\begin{equation*}
v_{t}=\frac{\pi eV}{4h_{c2}\left( t\right) T_{c0}},
\end{equation*}
which defines the characteristic scale of $\sigma^{\left( \mathrm{reg}\right)}$ in the considered domain of the phase diagram.
We stress, that this scale depends on temperature via the parameter $h_{c2}\left( t\right)$.

\emph{Close to $H_{c2}\left( 0\right)$, in the region of very low
temperatures $t\ll \widetilde{h}$}, the argument of the $\psi$-function in Eq.~(\ref{curr1a}) becomes large despite the smallness of $\widetilde{h}$, and the
$\psi$-function can therefore be approximated by its asymptotic expression.
One gets
\begin{align}
I^{(\mathrm{reg})}\left( v,t\ll \widetilde{h}\right) = -\frac{eS\Delta _{%
\mathrm{BCS}}}{4\pi ^{2}\sigma _{n}R_{N}}  \notag \\
\cdot \frac{v}{1+v^{2}}\left[ \ln \frac{\sqrt{1+v^{2}}}{%
\widetilde{h}}-\frac{\arctan v}{v}\right]  \label{QF}
\end{align}%
with $\Delta _{\mathrm{BCS}}=\pi T_{c0}/\gamma _{E}$ being the value of BCS\
gap. The characteristic scale where the maximum of the tunnel conductance
appears at these low temperatures is $v=2eV/\Delta _{\mathrm{BCS}}\sim 1$, i.e.
\begin{equation}
eV_{\max }\sim \Delta _{\mathrm{BCS}}.  \label{vmax}
\end{equation}

\emph{In the region of high fields $H\gg H_{c2}$ and low temperatures}, the
asymptotic behavior of the tunneling current can be studied in complete analogy to the case of high temperatures and weak fields.
The sums in Eq. (\ref{I1}) can be approximated by
integrals, which gives for the value of the differential conductivity at zero voltage:
\begin{equation*}
\sigma ^{(\mathrm{reg})}\left( 0,h\gg 1\right) =-\frac{e^{2}S}{4\pi
^{2}\sigma _{n}R_{N}}\left( \ln \frac{\ln \frac{1}{T_{c0}\tau }}{\ln h}%
\right)\,.
\end{equation*}%
One can see that this dependence is exactly the same as that one in the case of high temperatures with reversed roles of the reduced temperature and the reduced field.

\section{Low voltage behavior of $I^{\left( \mathrm{an}\right) }\left(
V\right) $}

\begin{figure}[tbh]
\begin{center}
\includegraphics[width=0.99\columnwidth ]{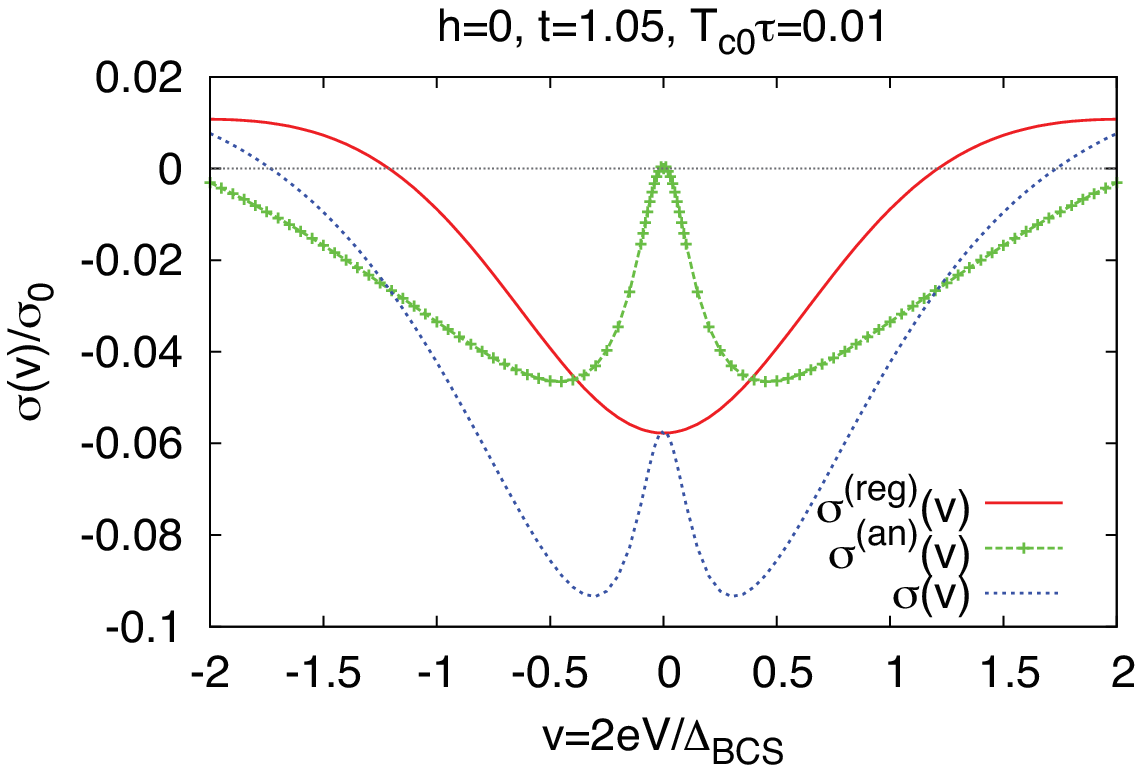} %
\includegraphics[width=0.99\columnwidth ]{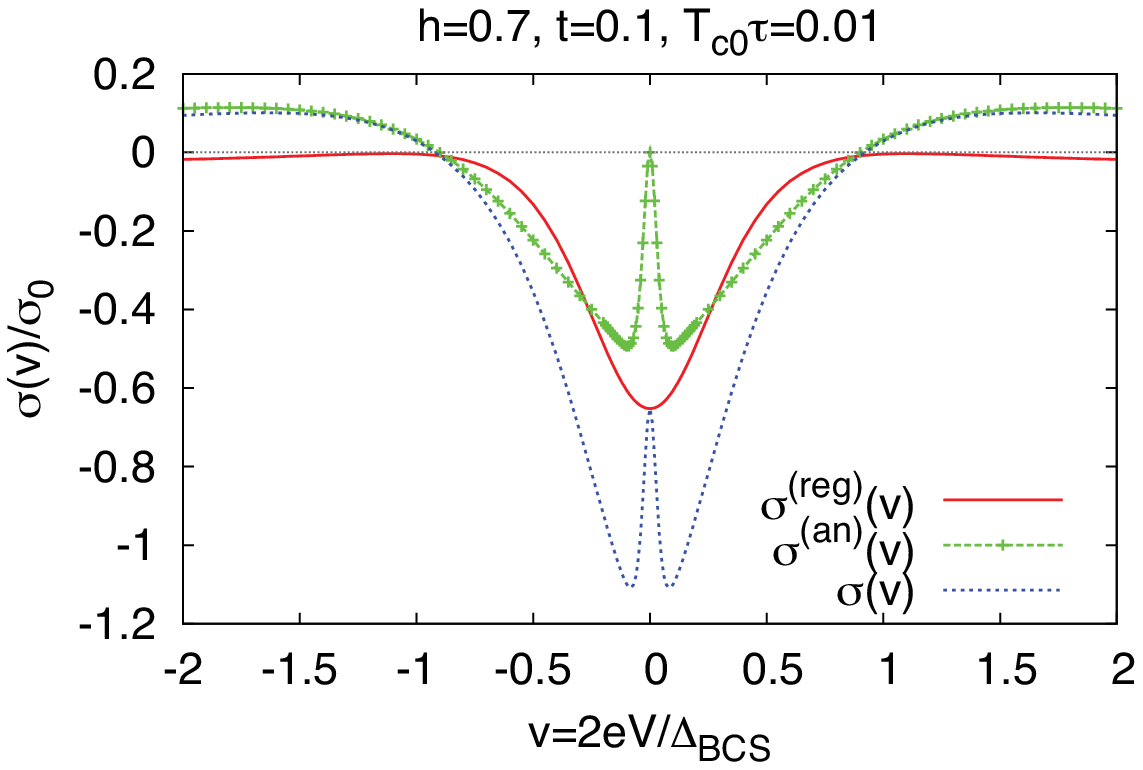}
\end{center}
\caption{(Color online) Regular and anomalous contributions to the tunneling conductance
close to $T_{c0}$ (top) and at low temperatures near $h_{c2}(0)$ (bottom). The regular part
is presented by a solid line (red), the anomalous by
crossed line (green) and the their sum, i.e., the total
fluctuation contribution, is shown by a dashed line (blue). (top) parameter point in Fig.~\ref{fig.regions} is $A_1$, (bottom) $B_1$.}
\label{tuncond_reg_an}
\end{figure}

In the low-voltage limit, $V\rightarrow 0$, the general expression for $I^{\left( \mathrm{an}\right) }\left( V\right)$, (\ref{Itot}), can be expanded
in small $eV$. We start with the first order term of that expansion, where one can assume $V=0$ in the argument of integrand function and
obtain

\begin{eqnarray}
&&I^{\left(  \mathrm{an}\right)  }\left(  V\rightarrow0\right)\!
=\!\frac{eT_{c0}Sh}{\pi^{3}\sigma_{n}R_{N}}\sum_{m=0}^{M}\left\{  \frac
{\Imm\Em_{m}^{\prime}\left( \! -ieV/\pi T\right)
}{\Em_{m}\left(  0\right)  }\right.\!+ \nonumber \\
&&\!\!\left.\frac{eV}{T}\int_{-\infty}^{\infty
}dz\frac{\Imm\Em_{m}\left(  iz\right)  \Imm%
\Em_{m}^{\prime}\left(  iz\right)  }{\sinh^{2}\pi z\left[
\Rem^{2}\Em_{m}\left(  iz\right) \! +\!\Imm%
^{2}\Em_{m}\left(  iz\right)  \right]  }\right\}.\label{ian}%
\end{eqnarray}

In the region of temperatures close to the transition temperature $T_{c0}$ and along the transition
line for temperatures $t\ll h_{c2}(t)$,  the propagator has a simple pole
structure [see Eqs.~(\ref{em0}) and (\ref{eline})] and the integral in Eq.~(\ref{ian}) can be calculated analytically.
Performing this integration one finds that the second term of Eq.~(\ref{ian}) exactly annihilates the linear part of the first term.
This fact justifies the static approximation (zero frequency bosonic mode) made in
Refs.~[\onlinecite{VD83,CCRV90,V93,R93,L10}].
Yet, this static approximation turns out to be valid only for very low voltages.
Expanding the integrand in  Eq.~(\ref{Itot}) to higher orders in voltage reveals an unexpected result.  One can
see that the voltage $V$ enters the integrand of  Eq.~(\ref{Itot}) in two different
places: in the argument of $\Em_{m}\left( iz-ieV\right) $ in the numerator
and in the argument of $\sinh \left( \pi z-eV/2T\right) $ in the denominator.
The expansion of $\Em_{m}\left( iz-ieV\right) $  results in the
appearance of a weakly voltage-dependent term of the order of $O\left( V^{3}/T_{c0}^{3}\right)$  in $I^{\left( \mathrm{an}\right) }$, while, as
one can easily verify, the expansion of $\sinh ^{-1}\left( \pi
z-eV/2T\right) $ after integration leads to a very singular
correction%
\begin{equation*}
I^{\left( \mathrm{an}\right) }\left( V\right) \propto \left( \frac{eV}{T_{c0}%
}\right) ^{3}\left\{
\begin{tabular}{ll}
$\epsilon ^{-2}$ & $eV\ll T-T_{c0}$ \\
$\widetilde{h}^{-2}$ & $eV\ll \Delta _{\mathrm{BCS}}\widetilde{h}$%
\end{tabular}%
\right. .
\end{equation*}%
At zero magnetic field, the dip in the corresponding tunnel conductivity rapidly
develops on the scale $eV\sim T-T_{c0}$, while at zero temperature close
to $H_{c2}\left( 0\right) $ this happens on the scale $eV\sim \Delta _{%
\mathrm{BCS}}\widetilde{h}.$

\begin{figure}[tbh]
\begin{center}
\includegraphics[width=0.99\columnwidth ]{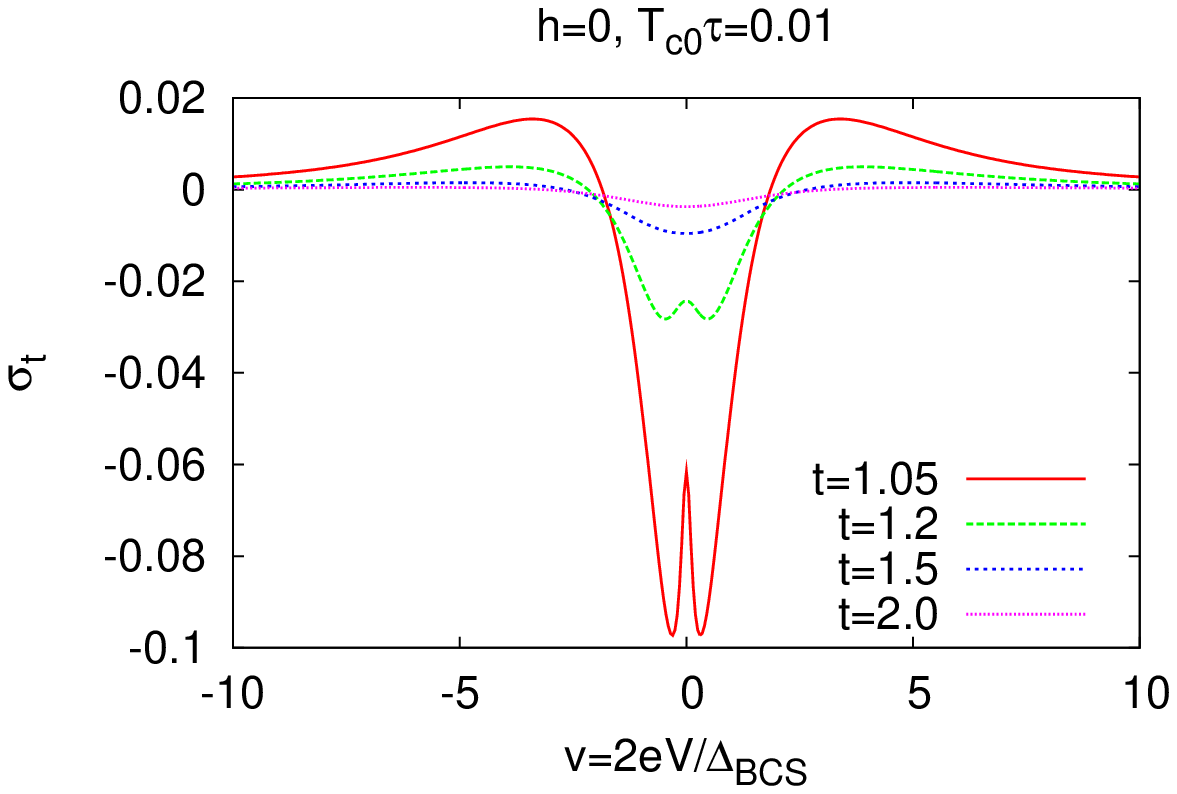}\\
\includegraphics[width=0.99\columnwidth ]{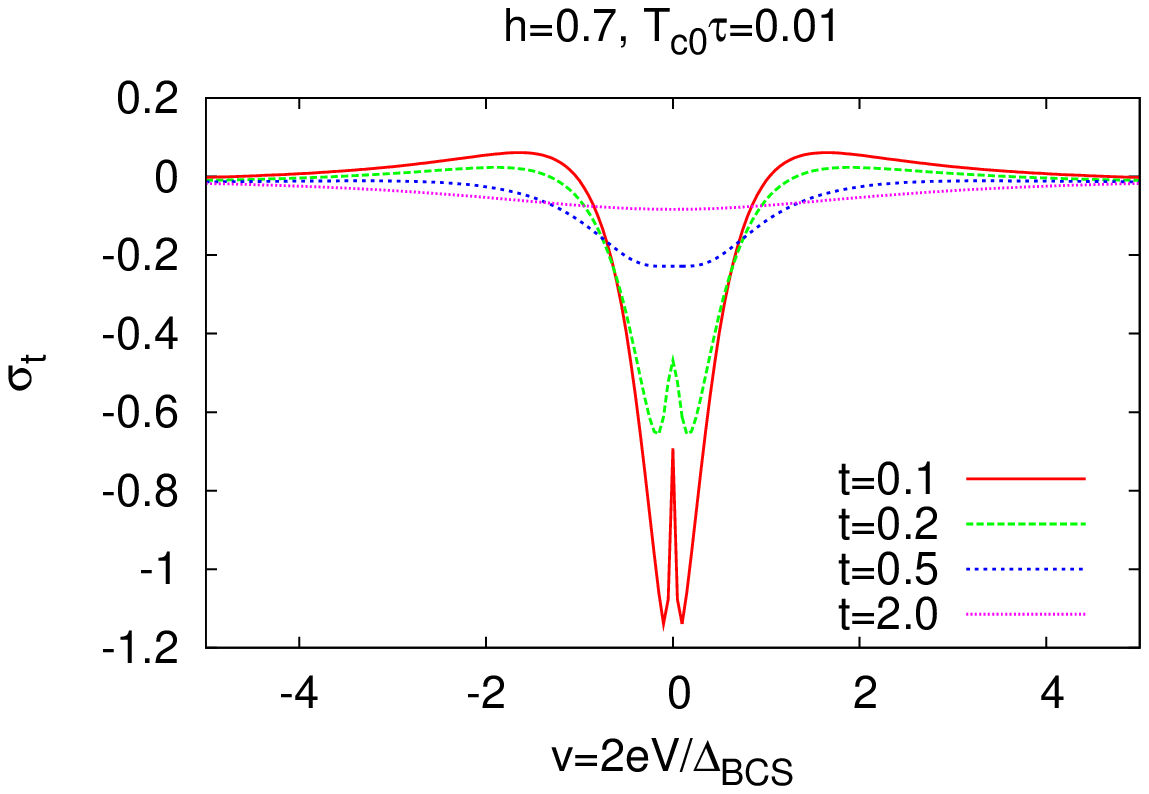}
\end{center}
\caption{(Color online) Smearing of the low-voltage singularity of the tunneling
conductance with increasing temperature close to $T_{c0}$ (top) and near $%
h_{c2}(0)$ (bottom). (top) parameter points in Fig.~\ref{fig.regions} are $A_1$-$A_4$,  (right) parameter points  are $B_1$-$B_4$.}
\label{singsmear}
\end{figure}

The effect of both fluctuation contributions, $I^{\left( \mathrm{reg}\right) }\left( V\right) $ and $%
I^{\left( \mathrm{an}\right) }\left( V\right)$, on the tunneling conductance is demonstrated in Fig.~\ref{tuncond_reg_an}.
Similar behavior can be observed along the whole line $H_{c2}\left( T\right) .$\
The singularity in the low voltage behavior of tunneling conductance rapidly
smears out when moving away from the transition line  or increasing the
temperature (see Fig.~\ref{singsmear}).

\section{Numerical analysis}

\begin{figure}[tbh]
\begin{center}
\includegraphics[ width=0.99\columnwidth ]{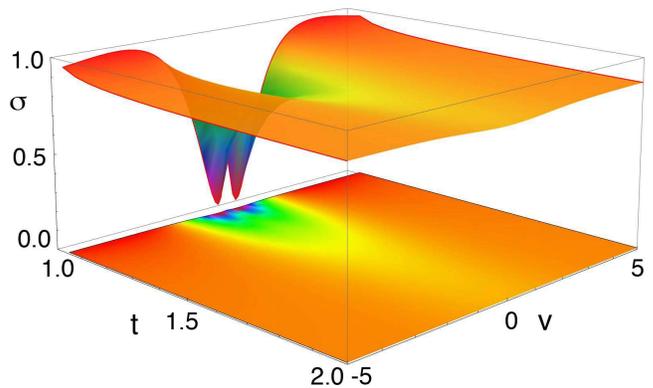}
\end{center}
\caption{(Color online) Temperature and voltage dependence of the tunneling conductance due
to superconducting fluctuations above $T_{c0}$ in zero magnetic field. The corresponding parameter line in Fig.~\ref{fig.regions} is $L_1$.}
\label{3DT}
\end{figure}

\begin{figure}[tbh]
\begin{center}
\includegraphics[ width=0.99\columnwidth]{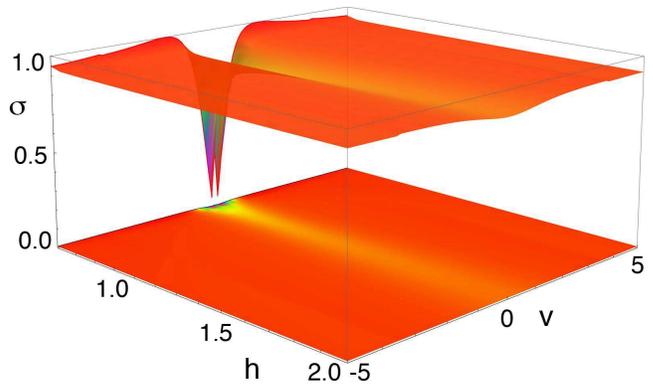}
\end{center}
\caption{(Color online) Magnetic field and voltage dependence of the tunneling conductance
due to superconducting fluctuations at $t=0.05$ above $h_{c2}(0)=0.69$. The corresponding parameter line in Fig.~\ref{fig.regions} is $L_2$.}
\label{3DH}
\end{figure}

The temperature, magnetic field, and voltage dependencies of the tunneling
conductance due to superconducting fluctuations, calculated numerically
based on Eq.~(\ref{Itot}), are presented in Figs.~\ref{3DT} - \ref{3DH} as surface plots.
The numerical procedure to calculate the $k$-sum of the first term needs to take into account its relatively slow convergence.
Therefore it is calculated explicitly up to a threshold at which the sum can be replaced by an integral and the polygamma functions by their asymptotic behavior.
(here we use as threshold-$k$, the value $k_M$ at which the argument of the function $\mathcal{E}_m$ reaches $1000$).
The ``rest''-integrals are calculated with inverse integration variable using a Gauss-Legendre method.
The second term requires a careful treatment of the two integrable poles, which is done by analytical calculation of the residuals in a small interval around them, where the denominator is linearized. Also the numerical integration outside the pole intervals is done by using adaptive integration point distances.
The overall behavior of both terms of the tunnel-current results in a
pronounced pseudo-gap structure of the conductance near the superconducting
region. It is the non-linear anomalous term of the tunnel-current which is
responsible for the fine structure (local maximum) at the center of the gap, the ZBA.

At this point it is worth mentioning that another sharp fine structure of tunnel conductance which should occur in the same scale $eV \sim T-T_{c0}$ was predicted in Ref.~[\onlinecite{VD83}]. This structure appears due to {\it interaction of fluctuations} as the second order correction in Ginzburg-Levanyuk number $\mathrm{Gi}_{\left(2\right)}$ (but still in first order in the barrier transparency).
This contribution has an interference nature (analogously to Maki-Thompson process) and, in contrast to the  discussed above nonlinear contribution $\sigma^{\left(\mathrm{an}\right) }\left( eV \ll T-T_{c0} \right)\sim \mathrm{Gi}_{\left(2\right)} [eV/(T-T_{c0})]^2$, diverges at zero voltage as $\mathrm{Gi}^2_{\left(2\right)}[T_{c0}/(T-T_{c0})]^2\ln [(T-T_{c0})/eV]$. Such divergency, in complete analogy to Maki-Thompson contribution, is cut off by any phase-breaking mechanism~\cite{M68,T70}.

Analyzing the surface plot representation of the experimental results of Ref.~[\onlinecite{Benjamen10}], obtained at temperature close to  $T_{c0}$, one notices their striking similarity to the theoretical surfaces presented in Fig.~\ref{3DT}.
Indeed, the authors of Ref.~[\onlinecite{Benjamen10}] mentioned the agreement of their results with the theoretical prediction of Ref.~[\onlinecite{VD83}].
Fig.~\ref{3DH} shows how the corresponding surface transforms at low temperatures and strong magnetic fields close to $H_{c2}(0)$.
\begin{figure}[tbh]
\begin{center}
\includegraphics[ width=0.99\columnwidth]{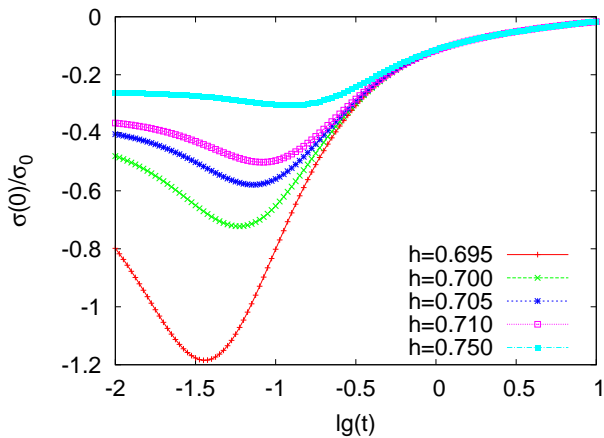}
\end{center}
\caption{(Color online) Zero-bias tunnel conductance, $\sigma(t,h,v=0)$, as function of temperature at fields slightly above the critical zero-temperature field for $h=0.695,0.7,0.705,0.71,0.75$. These correspond to the parameter line group $\mathrm{LL}_2$ in Fig.~\ref{fig.regions}.}
\label{fig.Tsig0_hc2}
\end{figure}

\begin{figure}[tbh]
\begin{center}
\includegraphics[ width=0.99\columnwidth]{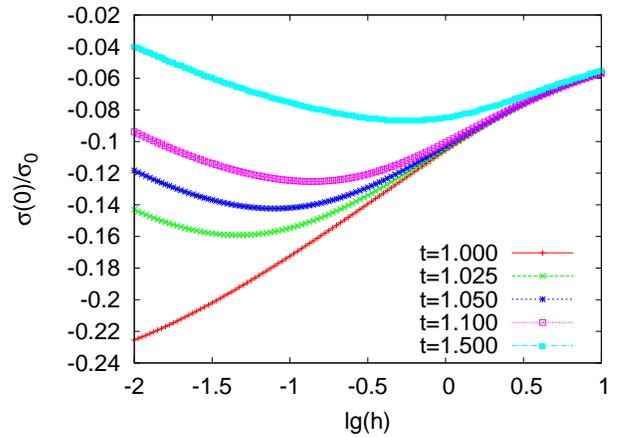}
\end{center}
\caption{(Color online) Zero-bias tunnel conductance as function of magnetic field near the critical temperature for $t=1.0,1.025,1.05,1.1,1.5$. These correspond to the parameter line group $\mathrm{LL}_1$ in Fig.~\ref{fig.regions}.}
\label{fig.Tsig0_tc}
\end{figure}

\begin{figure}[tbh]
\begin{center}
\includegraphics[ width=0.99\columnwidth]{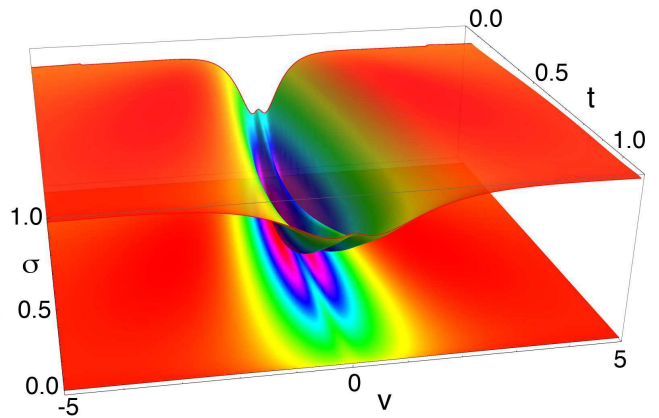}
\end{center}
\caption{(Color online) Voltage dependence of the tunneling conductance due to
superconducting fluctuations along the $h_{c2}(t)$ plotted for $(t,h)=s(x,h_{c2}(x))$ with $x\in [0;1]$ and distance $s=1.1$. The corresponding parameter line in Fig.~\ref{fig.regions} is $L_3$.}
\label{3DHC2}
\end{figure}

It is interesting to note that the behavior of the general expression~(\ref{Itot}) clearly shows growth of the fluctuation effects in the domain of intermediate temperatures and magnetic fields, beyond the immediate vicinity of $T_{c0}$ and $H_{c2}(0)$, see plots of the zero-bias tunnel conductance $\sigma(t,h,0)=\sigma^{\mathrm{(reg)}}(t,h,0)$ in Figs.~\ref{fig.Tsig0_hc2} and ~\ref{fig.Tsig0_tc}.  
In Fig.~\ref{3DHC2} one can see the evolution of the pseudogap near the $h_{c2}(t)$ line (slightly offset by a factor $1.1$, see caption), exhibiting a deeper suppression for intermediate temperatures and fields.
This fact is in agreement with the general ideas of the theory of fluctuations establishing the growth of fluctuations strength (characterized by the Ginzburg-Levanyuk number) as one moves away from the extreme points [$T_{c0}$ and $H_{c2}(0)$] of the curve $H_{c2}(T)$ (see chapter 2 of Ref.~[\onlinecite{LV05}]).

\section{Discussion and comparison with experiments}

\begin{figure}[tbh]
\begin{center}
\includegraphics[ width=0.99\columnwidth]{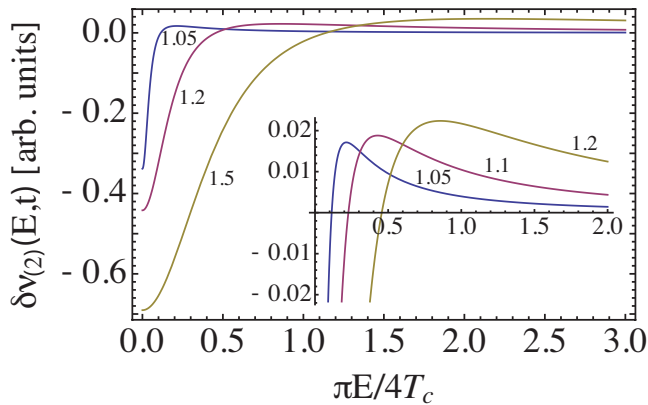}
\end{center}
\caption{(Color online) Theoretical curve  of the fluctuation correction to the single particle DOS, $\delta\nu_{(2)}^{\mathrm{(fl)}}$, versus energy, $E$, for 2D superconductors above critical temperature for temperatures $t=1.05,1.2,1.5,2.0$. The inset shows more details for temperatures close to $T_c$ near the maximum. The corresponding parameter points in Fig.~\ref{fig.regions} are $A_1$-$A_3$.}
\label{fig.dos}
\end{figure}

We have calculated the effect of SF on the tunnel-current characteristics for a SIN junction of a two-dimensional disordered superconductor and a normal metal electrode.
In contrast to the  common believe that the shape of $\sigma _{\mathrm{tun}}$, Eq.~(\ref{eq.sigmatun}), presents an almost direct image of the energy dependence of the density of states  of a measured  sample~\cite{density}, one notices, that this is not the case for the SIN system studied here.
Indeed, the effect of fluctuation Cooper pairing above $T_c$ results in the depletion of the electron density of states at the Fermi level ($E=0$) and its growth at the characteristic energies of the Cooper pairs, $E=T-T_c$. This phenomenon was considered theoretically long ago (see Ref.~[\onlinecite{ARW70,CCRV90}]) and the corresponding curves for $\delta \nu _{\left(2\right) }^{\left( \mathrm{fl}\right) }\left( E\right) $ in the absence of magnetic fields and not too far away from the Fermi level are presented in Fig.~\ref{fig.dos}. Comparing these curves to our $\sigma _{\mathrm{tun}}^{\left( \mathrm{fl}\right)}\left( V,T\right)$  (see Figs.~\ref{tuncond_reg_an} and \ref{singsmear}) reveals an apparent discrepancy, which is a result of the dramatic energy dependence of $\delta \nu _{\left(2\right) }^{\left( \mathrm{fl}\right) }\left( E\right) $ at small energies ($E \leq T$).

Note, that the statement concerning the similarity of the differential tunnel conductance versus voltage and the energy dependence of the density of states, is based on the common assumption that at low temperatures the latter changes on the scale $E \gg T$ and $\partial n_{F}/\partial E \sim \cosh^{-2}{E/2T}$ acts as a delta-function in Eq.~(\ref{eq.sigmatun}). 
This is exactly the case of the ZBA occurring in disordered metals~\cite{AA79}, where the correction to the density of states due to inter-electron interaction varies on the scale $E \sim \tau ^{-1} \gg T$ and $\delta \sigma _{\mathrm{tun}}^{\left( \mathrm{ZBA}\right)}\left( V\right) \sim \delta \nu ^{\left( \mathrm{Coulomb}\right) }\left(V\right)$.

However, the situation is just the opposite for the single-electron density of states which is renormalized by superconducting fluctuations.
Here the function $\delta \nu _{\left( 2\right)}^{\left( \mathrm{fl}\right) }\left( E,T\right)$ strongly varies and even changes its sign on the scale $E\sim T-T_{c0}\ll T$, while vanishing at large energies. 
In the energy interval, $E\in [0;\mathcal{O}(T)]$, the derivative $\partial n_{F}/\partial E$  remains practically equal to one.
Extending the  approximation $\partial n_F/\partial E=1$ to the complete integration domain $E\in [0;\infty)$, one would obtain zero due to the sum rule  $\int_{0}^{\infty }\delta \nu _{\left( 2\right) }^{\left( \mathrm{fl}%
\right) }\left( E,T\right) dE=0$ (the latter reflects the conservation of the number of states).
In order to obtain a finite conductivity, one needs to analyze Eq.~(\ref{eq.sigmatun}) more carefully, taking into account the high-energy tail
of $\delta \nu _{\left( 2\right)}^{\left( \mathrm{fl}\right) }\left( E,T\right) $, which is beyond the approximations used in Refs.~[\onlinecite{ARW70,CCRV90}]. 
Following the method proposed in Ref.~[\onlinecite{VD83}]), we calculated $\sigma _{\mathrm{tun}}^{\left( \mathrm{fl}\right)}\left( V,T\right) $ diagrammatically, without ever referring directly to the density of states, which consequently avoids the above mentioned limitations.
As a result of our detailed analysis of the tunnel-current, the wide bell-shaped form of the tunnel conductance with peripheral maximum at $eV_{\max }\sim k_{B}T_{c0}$ and sharp peak occurring at zero voltage, the ZBA, was found (see Figs.~\ref{tuncond_reg_an} and \ref{singsmear}).

In recent experimental works on Bi-$2212$  and Nb samples~\cite{K09, K11}, very interesting and carefully measured results on the temperature and magnetic field dependence of the differential tunnel conductivity are presented.
Unfortunately, a direct comparison of our results with those experiments is impossible at this point, since the authors of Refs.~\cite{K09, K11} investigated SIS junctions, while we consider SIN (or STM)  systems here.
Nevertheless, a careful study of Refs.~[\onlinecite{K09, K11}] reveals several similarities between the experimental finding and our prediction.
First, the authors always observe a peak in the fluctuation region ($T>T_c(H)$), which corresponds to energies smaller than $2\Delta_{BCS}(T=0)$ (Figs.~11 \& 12  of  Ref.~[\onlinecite{K09}]).
Secondly, very characteristic is the presence of a ``crossing point" in the temperature dependencies of the differential tunnel conductivity (see Figs. 4 \& 5b of Ref.~[\onlinecite{K09}]), analogous to that one of our Fig.~\ref{singsmear}.
Finally, the experiments also showed a very strong dependence of the tunnel magneto-conductivity on voltage close to $V=0$ (see Figures 6 d,c and 7 of Ref.~[\onlinecite{K11}]), which could be identified as the predicted {\it nonlinear anomalous contribution} [the second term of Eq.~(\ref{Itot}), which is rapidly smeared out when one moves away from the transition line or increases temperature (see Fig.~\ref{singsmear})].
However, in view of  the SIS-nature of the junctions used in the experiment, the observed ZBA could also be attributed to presence of a residual Josephson current. To the best of our knowledge this is the only case where the existence of a ZBA in the tunnel-conductivity close to the critical temperature is reported.  A definitive identification of its origin requires further studies.

The difficulty in observing the predicted ZBA due to $I^{(\textrm{an})}(V)$ can be related to the strong dependence of this ZBA on inhomogeneities in the barrier and local variations of the critical temperature.  Regrading the above mentioned second order correction due to the interaction of fluctuations, it was never reported to be observed experimentally.

\section{Acknowledgements}

The authors are grateful to T.~Baturina for attracting their interest to the problem of fluctuation formation of pseudogap close to $H_{c2}(0)$ and discussions.
We acknowledge useful discussions with B.~Altshuler, A.~Frydman, A~Goldman,  A.~Kamenev, V.~Krasnov, and M.~Norman.
The work was supported by the U.S. Department of Energy Office of Science under the Contract No. DE-AC02-06CH11357.
A.A.V. acknowledges support of the FP7-IRSES program, grant N 236947 ``SIMTECH''.


\begin{thebibliography}{99}
\bibitem{density}
We assume the density of states of one of the electrodes (e.g. the STM tip) to be constant.
It is well known that in disordered metals the Altshuler-Aronov zero bias anomaly (ZBA) of tunneling conductivity~\cite{AA79} appears due to the Coulomb interaction, $\delta\sigma_{\left( \mathrm{2}\right) }^{\left( \mathrm{ZBA}\right) }\left(E\right) \sim \ln\left( E\tau\right)$.
In that case, the electron scattering time $\tau^{-1}\gg T$ determines the characteristic energy scale of the density of states variation.
\end{thebibliography}
\end{document}